\documentclass[12pt]{article}
\usepackage{epsf,array}
\begin{document}
\begin{titlepage}
\rightline{QMW-PH-98-16}
\rightline{KCL-TH-98-13}
\def\today{\ifcase\month\or
        January\or February\or March\or April\or May\or June\or
        July\or August\or September\or October\or November\or December\fi,
  \number\year}
\rightline{hep-th/9803216}
\vskip 1cm
\centerline{\Large \bf  
Branes and Calibrated Geometries}
\vskip 0.8cm
\centerline{\sc Jerome P. Gauntlett$^1$, 
Neil D. Lambert$^2$ and Peter C. West$^2$}
\bigskip
\bigskip

\centerline {$^1${\it Physics Department}}
\centerline{{\it Queen Mary and Westfield College}}
\centerline{{\it Mile End Rd, London E1 4NS, U.K.}}
\centerline{{J.P.Gauntlett@qmw.ac.uk }} 
\bigskip

\centerline {$^2${\it Department of Mathematics}}
\centerline{{\it King's College}}
\centerline{{\it The Strand, London}}
\centerline{{\it WC2R 2LS, UK}}
\centerline{{ lambert, pwest@mth.kcl.ac.uk}}

\bigskip
 
\centerline{\sc Abstract}

The fivebrane worldvolume theory in eleven dimensions 
is known to contain BPS threebrane solitons which can
also be interpreted as a fivebrane whose worldvolume is
wrapped around a Riemann surface. 
By considering configurations of intersecting fivebranes and
hence intersecting threebrane solitons, we determine 
the Bogomol'nyi equations for more general BPS configurations. 
We obtain differential equations, generalising Cauchy-Riemann equations,
which imply that the worldvolume of the fivebrane is wrapped around
a calibrated geometry. 

\end{titlepage}
\newpage

\def\beq{\begin{equation}}
\def\eeq{\end{equation}}
\def\bea{\begin{eqnarray}}
\def\eea{\end{eqnarray}}
\renewcommand{\arraystretch}{1.5}
\def\ba{\begin{array}}
\def\ea{\end{array}}
\def\bce{\begin{center}}
\def\ece{\end{center}}
\def\nn{\noindent}
\def\nonu{\nonumber}
\def\pbx{\partial_x}

\font\mybb=msbm10 at 12pt
\def\bb#1{\hbox{\mybb#1}}
\def\bZ {\bb{Z}}
\def\bR {\bb{R}}
\def\bE {\bb{E}}
\def\bT {\bb{T}}
\def\bM {\bb{M}}
\def\bH {\bb{H}}
\def\hk {hyper-K{\" a}hler}
\def\HK {Hyper-K{\" a}hler}
\def\bfomeg{\omega\kern-7.0pt\omega}
\def\bfOmeg{\Omega\kern-8.0pt\Omega}

\def \Q {{\cal Q}}
\def \H {\td H}
\def \la{\longrightarrow}
\def \up {\uparrow}
 \def \upa {\uparrow}
 \def \nea {\nearrow}
\def \pa {\Vert}
\def\ma{\mapsto}
\def\inv{^{-1} }
\def \K {{\tilde K}}
\def\ww {\omega _{km } }
\def\mn {\mu\nu}

\def\np {{  Nucl. Phys. }}
\def\NP {{  Nucl. Phys. }}
\def \pl {{  Phys. Lett. }}
\def \PL {{  Phys. Lett. }}
\def \mpl {{ Mod. Phys. Lett. }}
\def \prl {{  Phys. Rev. Lett. }}
\def \PRL {{  Phys. Rev. Lett. }}
\def \pr  {{ Phys. Rev. }}
\def \PR  {{ Phys. Rev. }}
\def \ap  {{ Ann. Phys. }}
\def \cmp {{ Commun.Math.Phys. }}
\def \CMP {{ Commun.Math.Phys. }}
\def \ijmp {{ Int. J. Mod. Phys. }}
\def \jmp {{ J. Math. Phys.}}
\def \cqg {{ Class. Quant. Grav. }}
\def \CQG {{ Class. Quant. Grav. }}


\def\p{\partial}
\def\del{\partial }
\def\a {\alpha}
\def\b{\beta}
\def\g{\gamma} 
\def\Ga{\Gamma}
\def\de{\delta} \def\De{\Delta}
\def\e{\epsilon}
\def\vep{\varepsilon}
\def\ze{\zeta}
\def\et{\eta}
\def\t{\theta} \def\Th{\Theta}
\def\vth{\vartheta}
\def\io{\iota}
\def\ka{\kappa}
\def\l{\lambda} 
\def\La{\Lambda}
\def\rh{\rho}
\def\s{\sigma} \def\Si{\Sigma}
\def\ta{\tau}
\def\up{\upsilon} 
\def\Up{\Upsilon}
\def\Ph{\Phi}
\def\vph{\varphi}
\def\ch{\chi}
\def\ps{\psi} 
\def\Ps{\Psi}
\def\om{\omega} 
\def\Om{\Omega}

\def \la{\longrightarrow}
\def\ha { {\textstyle {1\over 2}} }

\def\vol#1{{\bf #1}}
\def\nupha#1{Nucl. Phys., \vol{#1} }
 \def\CMP#1{Comm. Math. Phys., \vol{#1} }
\def\phlta#1{Phys. Lett., \vol{#1} }
\def\phyrv#1{Phys. Rev., \vol{#1} }
\def\PRL#1{Phys. Rev. Lett., \vol{#1} }
\def\prs#1{Proc. Roc. Soc., \vol{#1} }
\def\PTP#1{Prog. Theo. Phys., \vol{#1} }
\def\SJNP#1{Sov. J. Nucl. Phys., \vol{#1} }
\def\TMP#1{Theor. Math. Phys., \vol{#1} }
\def\ANNPHY#1{Annals of Phys., \vol{#1} }
\def\PNAS#1{Proc. Natl. Acad. Sci. USA, \vol{#1} }

\def\mapa#1{\smash{\mathop{\longrightarrow }\limits_{#1} }}
\def\x {\times}

\section{Introduction}

The dynamics of branes have played an important role in elucidating the
structure of M-theory
(for a review see~\cite{PKT}). 
In particular the fivebrane has received substantial interest recently 
due to its intricate worldvolume theory. This theory has been shown to contain 
supersymmetric threebrane~\cite{three} and self-dual string~\cite{one} 
solitons. A remarkable feature of these solitons, and closely related 
solitons on the worldvolumes of D-branes, is that they incorporate their 
spacetime  
interpretation~\cite{three,one,cm,gwg,BGT,GGT}. 
{}For example, the self-dual string  corresponds to a 
membrane ending on the fivebrane. Similarly, the simplest threebrane soliton
solution can be interpreted as the orthogonal intersection of two 
fivebranes lying along flat hyperplanes. In fact, for this case
the Bogomol'nyi equations are precisely the Cauchy-Riemann equations. Thus
there are more general solutions corresponding to
desingular deformations of this configuration which can be interpreted as a 
single fivebrane with its worldvolume wrapped around an arbitrary 
Riemann surface. 

There are solutions of the supergravity equations of motion 
corresponding to orthogonal intersections of branes, 
but the BPS solutions that are known at present 
are not fully localised~\cite{paptown,tseytlin,gkt} (for a review 
see~\cite{gauntrev}). 
The description of intersecting branes given by examining the worldsheet theory
thus provides a useful avenue of obtaining more insights into the properties 
of M-branes.
Moreover, the existence of branes with non-trivial worldvolumes
has important applications in relation to the low energy
dynamics of quantum Yang-Mills theories, e.g. the derivation of the 
Seiberg-Witten curve~\cite{Witten} (see also~\cite{KLVW})
and indeed all of the Seiberg-Witten  
dynamics~\cite{HLW} from the fivebrane.

It is natural to enquire if there are other BPS solutions of the 
worldvolume that correspond to intersecting threebranes and self-dual strings. 
{}From the supergravity point of view this seems rather natural: 
supersymmetric  configurations of orthogonal
intersecting membranes and fivebranes are known and we might expect to see 
analogous configurations in the worldvolume theory. {}For example,
a supersymmetric configuration is given by a fivebrane in 
$\{x^1,x^2,x^3,x^4,x^5\}$
plane orthogonally intersecting another fivebrane in the 
$\{x^3,x^4,x^5,x^6,x^7\}$ plane, with a membrane in the $\{x^3,x^6\}$ plane,
a configuration that we will denote 
\bea
\matrix{
M5:&1&2&3&4&5& & &\cr
M5:& & &3&4&5&6&7&\cr
M2:& &2& & & &6& &\cr}
\label{eq:twotwo}
\eea
Considering the first fivebrane's worldvolume theory we expect this
configuration to correspond, in the simplest setting,
to a BPS solution consisting of a threebrane soliton in the $x^3,x^4,x^5$ 
directions
orthogonally intersecting a self-dual string in the $x^2$ direction. 
This self-dual string then acts as a source for the three form field $h$
on the fivebrane worldvolume.
More general solutions should correspond to BPS solitons
in the fivebrane approach to N=2 superYang-Mills 
theory~\cite{Witten,henningsonyi}.

As a first step towards studying all supersymmetric configurations of branes, 
in this paper we will consider configurations with only fivebranes.
In the simplest setting these should correspond to intersecting configurations 
of threebranes on the worldvolume, but more generally they 
can be interpreted as the worldvolume
of a single fivebrane with a non-trivial worldvolume, i.e. these BPS states
may simply be viewed as a single fivebrane wrapped on a non-trivial 
submanifold embedded in eleven dimensions. Since there are no membranes and
we are considering solitons with only scalars active, 
our discussion is universal 
to all types of branes by dimensional reduction and T-duality. 
The fivebrane in eleven dimensions is particularly useful in this sense
because it has both a large worldvolume and transverse space. 
We will address the issue of configurations involving fivebranes, 
membranes and momentum modes in a future paper.
In our analysis we will choose the    
target space to be flat space throughout, although the
generalisation to a curved space should be straightforward and will be 
briefly discussed in the conclusion.

The supersymmetry of (Euclidean) 
membranes wrapped on three cycles of a Calabi-Yau
manifold and threebranes 
wrapped around three cycles and four cycles of exceptional holonomy 
manifolds has been  studied in~\cite{BBS,BBMOOY}.
{}From those results we expect the supersymmetric configurations 
of fivebranes to correspond 
to calibrated surfaces. In this work we shall focus on a  full description of 
the 
non-linear worldvolume theory of the fivebrane and its supersymmetry.  
In this way we hope to obtain a more detailed
picture. In particular our derivation shows that 
such surfaces satisfy elegant differential equations, generalising 
Cauchy-Riemann
equations, which 
appear in the work of Harvey
and Lawson~\cite{HL} as necessary and sufficient conditions for the 
manifold to be calibrated. In addition, since we will directly show
that the surfaces must be calibrated using similar ideas 
to~\cite{BBS,BBMOOY},
our results can be viewed as a supersymmetric proof of
some of the results in~\cite{HL}.

The plan of the rest of the paper is as follows.
In the next section we obtain a list of orthogonally intersecting 
fivebranes which
preserve some fraction of eleven-dimensional spacetime
supersymmetry. The purpose of this section is to characterise some 
features of potential
supersymmetric solutions on the fivebrane. In particular we will
identify which transverse scalars we expect to be active in the solutions
and determine sets of projection
operators acting on the supersymmetry parameters that will be useful 
in later sections.  
{}Following that we turn our 
attention to the non-linear
worldvolume theory of the fivebrane in section three.
{}For the reader who is not interested in all the details of this section,
we point them to equation (\ref{eq:susy}), 
which is the condition for the fivebrane
to preserve some supersymmetry in cases where 
the self-dual three form vanishes. 
{}Following this equation we present
the argument that the fivebranes must be wrapped along calibrated
geometries. 
In the section four we combine the results of sections two and three to
derive the Bogomol'nyi equations for
supersymmetric fivebrane
configurations.

\section{Intersecting Fivebranes}

In this section we construct a  number  of orthogonally
intersecting fivebrane configurations which preserve some fraction of
eleven-dimensional spacetime supersymmetry (see also~\cite{Groningen}) 
and list the corresponding supersymmetry projectors. This will provide a guide
in our search for Bogomol'nyi
conditions for supersymmetric solutions in the fivebrane worldvolume theory.
We first note that a fivebrane in the $\{x^0,x^1,x^2,x^3,x^4,x^5\}$
plane preserves the supersymmetries $\e\Gamma^{012345}=\e$, where
$\Gamma$ are the eleven-dimensional 
${\Gamma}$-matrices. The addition
of other fivebranes will therefore imply further projections on $\e$.
We shall list the various configurations in the order of the amount of 
supersymmetry that they preserve.

It turns out that in many configurations the supersymmetry conditions 
allow for additional fivebranes to be included, without breaking more 
supersymmetries. Thus the number of fivebranes can be rather large and does
not immediately reflect the amount of supersymmetry preserved. 
We follow the practice of always including these
extra fivebranes, which make the configurations more symmetric. However
we only list an independent set of projectors for each configuration.

The reader will note in the following that there is clearly some 
choice between adding fivebranes or anti-fivebranes, although only for
those fivebranes corresponding to independent projectors. Once these
fivebranes are fixed, there is no choice for the others. 
In this section however, we merely
wish to motivate the choice of projections used the in worldvolume analysis
in the following sections.
Clearly one could find other solitons by changing fivebranes
to anti-fivebranes and visa-versa. 
However this would only lead to trivial
changes in our analysis and correspond to changing the signs of the 
coordinates. 

\subsection{1/4 Supersymmetry}

\bea
\matrix{
M5:&1&2&3&4&5& & &\cr
M5:& & &3&4&5&6&7&\cr}
\label{eq:two}
\eea
\bea
\epsilon\Gamma^{012345}=\epsilon\ , \quad
\epsilon\Gamma^{1267}=-\epsilon\ .
\eea
This spacetime configuration should manifest itself as two active scalars 
($X^6,X^7$)
depending on two
worldvolume coordinates ($x^1,x^2$), i.e. a two-dimensional 
surface embedded in four dimensions. As mentioned above the differential 
equation that
the scalars satisfy in BPS solutions are simply Cauchy-Riemann equations, and 
hence
this situation corresponds to a fivebrane wrapped around a  
Riemann surface. 

\subsection{1/8 Supersymmetry}

\bea
\matrix{
M5:&1&2&3&4&5& & & & &\cr
M5:& & &3&4&5&6&7& & &\cr
M5:& & &3&4&5& & &8&9&\cr}
\label{eq:three}
\eea
\bea
\epsilon\Gamma^{012345}=\epsilon\ ,\quad
\epsilon\Gamma^{1267}=-\epsilon\ ,\quad 
\epsilon\Gamma^{1289}=-\epsilon\ .
\label{eq:threesusy}
\eea
BPS worldvolume solutions corresponding to this configuration should
have four active scalars depending on two worldvolume coordinates.
Thus it should appear as a two-dimensional surface embedded in six dimensions
(and moreover it must not be possible to embed the surface in four dimensions). 
In fact
it corresponds to a Riemann surface but this time embedded in a six dimensional
space. We note that one also has $\e\Gamma^{10}=-\e$.

\bea
\matrix{
M5:&1&2&3&4&5& & &\cr
M5:& & &3&4&5&6&7&\cr
M5:&1&2& & &5&6&7&\cr}
\label{eq:five}
\eea
\bea
\epsilon\Gamma^{012345}=\epsilon\ ,\quad
\epsilon\Gamma^{1267}=-\epsilon\ ,\quad 
\epsilon\Gamma^{3467}=-\epsilon\ .
\label{eq:fivesusy}
\eea
{}For this case we expect two active scalars depending on four worldsurface 
coordinates.
BPS solutions should  appear as a four surface embedded in six dimensions
and in fact corresponds to a complex manifold. Note that $\e\Gamma^{05}=-\e$ so
that we could add a pp-wave in the $x^5$ direction without breaking any more
supersymmetries.

\bea
\matrix{
M5:&1&2&3&4&5& & & &\cr
\overline{M5}:& & &3&4&5&6&7& &\cr
M5:& &2& &4&5&6& &8&\cr
\overline{M5}:&1& & &4&5& &7&8&\cr}
\label{eq:four}
\eea
\bea
\epsilon\Gamma^{012345}=\epsilon\ ,\quad
\epsilon\Gamma^{1267}=\epsilon\ ,\quad 
\epsilon\Gamma^{1368}=\epsilon\ .
\label{eq:foursusy}
\eea
This configuration should correspond to solutions with 
three active scalars depending on three worldvolume
coordinates. We will see that this corresponds to a three-dimensional 
special Lagrangian manifold embedded in six dimensions.

\subsection{1/16 Supersymmetry}

\bea
\matrix{
           M5:&1&2&3&4&5& & & & &\cr
           M5:&1& & &4&5&6& & &9&\cr
\overline{M5}:&1& & &4&5& &7&8& &\cr
\overline{M5}:&1&2& & &5& & &8&9&\cr
           M5:&1&2& & &5&6&7& & &\cr
           M5:&1& &3& &5&6& &8& &\cr
           M5:&1& &3& &5& &7& &9&\cr}
\label{eq:six}
\eea
\bea
\epsilon\Gamma^{012345}=\epsilon\ ,\quad
\e\Gamma^{2369}=-\e\ ,\quad \e\Gamma^{2378}=\e\ ,
\quad \e\Gamma^{3489}=\e\ .
\label{eq:sixsusy}
\eea
{}For this configuration  we should have four scalars depending on three 
worldvolume coordinates.
We will see below that it describes an associative three surface in 
seven dimensions. Note that we also have $\e\Gamma^{10}=\e$
for this configuration. 

\bea
\matrix{
           M5:&1&2&3&4&5& & & & &\cr
           M5:& & &3&4&5& & &8&9&\cr
           M5:& &2& &4&5& &7& &9&\cr
\overline{M5}:&1& & &4&5& &7&8& &\cr
\overline{M5}:&1&2& & &5& & &8&9&\cr
           M5:& &2&3& &5& &7&8& &\cr
           M5:&1& &3& &5& &7& &9&\cr}
\label{eq:seven}
\eea
\bea
\epsilon\Gamma^{012345}=\epsilon\ ,\quad
\e \Gamma^{1289}=-\e\ ,\quad \Gamma^{1379}=\e\ ,\quad 
\e \Gamma^{2378}=\e\ .
\label{eq:sevensusy}
\eea
Here we  should look for solutions with 
three scalars depending on four worldvolume coordinates. We will
see below that this corresponds to a coassociative four surface in 
seven-dimensions. 
Note that we have $\e\Gamma^{05}=\e$ so that we could add
a pp-wave in the $x^5$ direction without  breaking any more supersymmetries.

\bea
\matrix{
M5:&1&2&3&4&5& & & & &\cr
M5:& & &3&4&5&6&7& & &\cr
M5:&1&2& & &5& & &8&9&\cr
M5:& & &3&4&5& & &8&9&\cr
M5:&1&2& & &5&6&7& & &\cr
M5:& & & & &5&6&7&8&9&\cr}
\label{eq:eight}
\eea
\bea
\epsilon\Gamma^{012345}=\epsilon\ ,\quad
\epsilon\Gamma^{1267}=-\epsilon\ ,\quad 
\epsilon\Gamma^{3489}=-\epsilon\ ,
\quad \epsilon\Gamma^{1289}=-\epsilon\ .
\label{eq:eightsusy}
\eea
This configuration corresponds to four scalars depending on four worldvolume 
coordinates.
We will see that it corresponds to a complex four dimensional surface 
embedded in eight dimensions. Note that $\e\Gamma^{05}=-\e$, 
$\e\Gamma^{10}=-\e$ and 
$\e\Gamma^{0510}=\e$ so we could
add a pp-wave in the $x^5$ direction and 
a membrane in the $\{x^0,x^5,x^{10}\}$ plane. The presence of the membrane 
is related to the
fact that the second and third fivebranes intersect 
over a string, rather than a 
threebrane. We have not considered this string soliton by itself because
there is no known worldvolume solution to describe it. Such configurations
will appear again but unlike this case, where the orthogonal intersection is 
necessary
to obtain the corresponding projections, the 
fivebranes which contribute string intersections could be discarded.

\bea
\matrix{
M5:&1&2&3&4&5& & & & &\cr
\overline{M5}:& & &3&4&5&6&7& & &\cr
M5:& &2& &4&5&6& &8& &\cr
\overline{M5}:& &2&3& &5&6& & &9&\cr
\overline{M5}:&1& &   &4&5& &7&8& &\cr
M5:& & & & &5&6&7&8&9&\cr
M5:&1& &3& &5& &7& &9&\cr
\overline{M5}:&1&2& & &5& & &8&9&\cr}
\label{eq:fifteen}
\eea
\bea
\epsilon\Gamma^{012345}=\epsilon\ ,\quad
\epsilon\Gamma^{1267}=\epsilon\ ,\quad \epsilon\Gamma^{1368}= \epsilon\ ,
\quad \epsilon\Gamma^{1469}=\epsilon\ .
\label{eq:fifteensusy}
\eea
Here we again have four scalars depending on four worldvolume coordinates.
We will see below that this corresponds to a four-dimensional
special Lagrangian surface embedded in eight dimensions. Note that we also
have $\e\Gamma^{0510}=\e$ so again we could add a membrane in the 
$\{x^0,x^5,x^{10}\}$ plane.

\subsection{1/32 Supersymmetry}

\bea
\matrix{
           M5:&1&2&3&4&5& & & & &\cr
\overline{M5}:& & &3&4&5&6&7& & &\cr
           M5:& &2& &4&5&6& &8& &\cr
           M5:&1&2& & &5&6&7& & &\cr
           M5:&1& &3& &5&6& &8& &\cr
           M5:& &2&3& &5& &7&8& &\cr
\overline{M5}:&1& & &4&5& &7&8& &\cr
           M5:& & & & &5&6&7&8&9&\cr
\overline{M5}:& &2&3& &5&6& & &9&\cr
           M5:& & &3&4&5& & &8&9&\cr
           M5:& &2& &4&5& &7& &9&\cr
           M5:&1& &3& &5& &7& &9&\cr
           M5:&1& & &4&5&6& & &9&\cr
\overline{M5}:&1&2& & &5& & &8&9&\cr}
\label{eq:nine}
\eea
\bea
\epsilon\Gamma^{012345}=\epsilon\ ,\quad
\e \Gamma^{1267}=\e,\quad
\e \Gamma^{1368}=\e,\quad \Gamma^{1469} =\e\ , 
\quad\Gamma^{1289}=-\e.
\label{eq:nineproj}
\eea
In this configuration we expect four scalars depending on four worldvolume
coordinates. We will see below that this solution is described by a 
Cayley four surface in eight dimensions. Note that here
we have $\e\Gamma^{0510}=-\e$, $\e\Gamma^{05}=-\e$ and $\e\Gamma^{10} = \e$.
Thus we could add membranes in the $\{x^0,x^5,x^{10}\}$ plane and 
pp-waves in the $x^5$ direction
without breaking any additional
supersymmetry.

\bea
\matrix{
M5:&1&2&3&4&5& & & & &  &\cr
\overline{M5}:& & &3&4&5&6&7& & &  &\cr
M5:& &2& &4&5&6& &8& &  &\cr
\overline{M5}:& &2&3& &5&6& & &9&  &\cr
M5:& &2&3&4& &6& & & &10&\cr
\overline{M5}:&1& & &4&5& &7&8& & &\cr
M5:&1& &3& &5& &7& &9&  &\cr
\overline{M5}:&1&2& & &5& & &8&9&  &\cr
\overline{M5}:&1& &3&4& & &7& & &10&\cr
M5:&1&2& &4& & & &8& &10&\cr
\overline{M5}:&1&2&3& & & & & &9&10&\cr
M5 :& & & & &5&6&7&8&9&  &\cr
\overline{M5}:& & & &4& &6&7&8& &10&\cr
M5:& & &3& & &6&7& &9&10&\cr
\overline{M5}:& &2& & & &6& &8&9&10&\cr
M5:&1& & & & & &7&8&9&10&\cr}
\label{eq:sixteen}
\eea
\bea
\epsilon\Gamma^{012345}=\epsilon\ ,\quad
\epsilon\Gamma^{1267}=\epsilon\ ,\quad \epsilon\Gamma^{1368}= \epsilon\ ,
\quad \epsilon\Gamma^{1469}= \epsilon\ ,
\quad \epsilon\Gamma^{15610}= \epsilon\ .
\label{eq:sixteensusy}
\eea
In this configuration all five scalars are active and depend on all five
worldvolume coordinates. We will see that it manifests itself as
a five-dimensional special Lagrangian surface in ten dimensions.
Again there are fivebranes intersecting over
strings and
$\e\Gamma^{0510}=\e$, $\e\Gamma^{049}=\e$, 
$\e\Gamma^{038}=\e$, $\e\Gamma^{027}=\e$ and $\e\Gamma^{016}=\e$
so that we can add membranes in the $\{x^0,x^1,x^6\}$, $\{x^0,x^2,x^7\}$,  
$\{x^0,x^3,x^8\}$, $\{x^0,x^4,x^9\}$ and $\{x^0,x^5,x^{10}\}$ planes.

\section{Supersymmetry and the Fivebrane}

In this paper we  are interested in bosonic solutions of the fivebrane equations 
of 
motion that preserve some supersymmetry. This will
be the case if there exists constant spinors such that the
variation of the spinor field of the fivebrane theory
vanishes: the resulting condition is the Bogomol'nyi equation for the bosonic 
fields. 
We will see that the Bogomol'nyi condition will determine the 
geometry of the fivebrane configuration. In this section we derive an explicit 
expression for the supersymmetric variation of the spinor field of the 
fivebrane for the case of vanishing self-dual three form, 
generalising and refining the discussion found in~\cite{one}.

We use the
fivebrane dynamics and conventions of~\cite{HSW}. In our paper the
fivebrane is embedded in flat eleven-dimensional  Minkowski
superspace. We must distinguish between world and tangent indices,
fermionic and bosonic indices  and indices associated with the target
space
$\underline M$ and the fivebrane worldvolume $M$. On the fivebrane
worldvolume   the bosonic tangent space indices are denoted by
$a,b,...= 0,1,2,...,5$ and bosonic world indices by
$m,n,...=0,1,2,...,5$. {}For example,  the inverse
vielbein of the bosonic sector of the fivebrane worldvolume is denoted by
$E_a^{\ m}$. The bosonic indices of the tangent space of the
target space $\underline M$ are denoted by the same symbols, but
underlined, i.e. the inverse vielbein in the bosonic sector is given
by $E_{\underline a}^{\ \underline m}$. The fermionic indices follow the
same pattern,  those in the tangent space are denoted by $\alpha $
and $\underline \alpha$ for worldvolume $M$ and target space
$\underline M$
respectively, while the world spinor indices are denoted by $\mu$ and
$\underline \mu$.

The fivebrane sweeps out a superspace $M$ in the target superspace
$\underline M$
which is specified in local coordinates
$Z^{\underline M}= (X^{\underline m}, \Theta ^{\underline \mu})
,\ \underline m=0,1,\dots ,10,\ \underline \mu=1,\dots,32$.
These coordinates are functions of the worldvolume superspace
parameterised by $z^{ M}= (x^{ m}, \theta^{ \mu}),\  m=0,1,\ldots,5;
\ \mu = 1,\ldots ,16$. The $\theta^{ \mu}$ expansion
of the $Z^{\underline M}$ contains
$x^m$ dependent fields of which the only independent ones  are their
$\theta^\mu=0$ components, also denoted
$X^{\underline m}$ and $ \Theta ^{\underline \mu}$, and a
self-dual tensor $h_{abc}$ which occurs at level $\theta ^\mu $
in $ \Theta ^{\underline \mu}$. Despite the redundancy of notation
it  will be clear from the context when we are discussing the component
fields and the superfields.

The bosonic target space indices tangent to $\underline M$ may be decomposed as
those that lie in the  fivebrane worldvolume and those that lie in the
space transverse to the fivebrane; we denote these indices by $a$ and
$a'$ respectively  (i.e. $\underline a=(a,a'), a=0,1,\ldots,5;\
a'=1',\ldots,5'$)\footnote{We will also use $a'=6,7,8,9,10$.} 
with a similar convention for  world indices. The
initially thirty-two component spinor indices $\underline \alpha$ are
split into a pair of sixteen component spinor indices (i.e.
$\underline \alpha=(\alpha,
\alpha'),\ \alpha=1,\ldots,16;\ \alpha'=1',\ldots,16'$) corresponding to
the breaking of half of the supersymmetries by the fivebrane.

We will use the super-reparameterisations of the
worldvolume to choose the so-called  static
gauge. In this gauge we identify the bosonic  coordinates in
the worldvolume with the bosonic   coordinates on  the worldvolume
(i.e. $ X^n=x^n,\ n=0,1,\ldots ,5$) and set the fermionic fields
$\Theta ^\alpha=0, \  \alpha=1,\ldots,16$. {}For a flat background
$\Theta ^{\underline \mu}=\Theta ^{\underline \alpha}
\delta _{\underline \alpha}^{\underline \mu}$. The component
field content of the fivebrane is
$X^{a'}$ $(a'=1',\dots,5')$, $\Theta ^{\alpha'}$ $(\alpha'=1',\ldots,16')$
and the self-dual field strength
$h_{abc}$.

We  recall some of the salient points
of the super-embedding formalism. The frame vector fields on the target
manifold $\underline M$ and the fivebrane worldvolume submanifold
$M$ are given by $E_{\underline A} = E_{\underline A} ^{\ \underline M}
\partial_{\underline M}$ and
$E_{A} = E_{A} ^{\ M} \partial_{M}$ respectively.
The coefficients $E_{A}^{\ \underline A}$ encode the relationship
between the vector fields $E_{A}$ and $E_{\underline A}$, i.e. $E_{A}
= E_{A}^{\ \underline A}E_{\underline A}$. Applying this relationship
to the coordinate $Z^{\underline M}$ we find the equation
\bea
E_{A}^{\ \underline A}= E_{A}^{\ \ N}\partial_N Z^{\underline M}
E^{\ \underline A}_{\underline M}\ .
\label{eq:susytwo}
\eea

In this paper we will be primarily interested in fivebranes whose
worldvolumes have $h_{abc}=0$. In this case the geometry of the
fivebrane simplifies considerably. The vector fields
$E_{\overline \alpha}^{\
\underline \beta}\equiv (E_{\alpha}^{\ \underline \beta},
E_{\alpha'}^{\ \underline \beta})$ and 
$E_{\overline a}^{\ \underline a}\equiv (E_{a}^{\ \underline b},
E_{a'}^{\ \underline b})$ on the
fivebrane can be chosen to be equal to the $Spin(1,10)$ and $SO(1,10)$
matrices $u_{\overline \alpha}^{\
\underline \beta}$and $u_{\overline a}^{\ \underline a}$ respectively.
{}For example
\bea
E_{\alpha}^{\ \underline \beta} = u_{\alpha}^{\ \underline
\beta} , \quad
 E_{\alpha'}^{\ \underline \beta} = u_{\alpha'}^{\ \underline \beta}\ ,\quad
E_{a'}^{\ \underline a} =  u_{a'}^{\ \underline a}\ .
\label{eq:susyfour}
\eea
The matrix $u_{\overline a}^{\ \underline b}
\equiv (u_{a}^{\  b},u_{a'}^{\ b'})$ is an  element of 
$SO(1,10)$ and the matrix
$u_{\overline \alpha}^{\ \underline \beta}\equiv
(u_{\alpha}^{\ \underline \beta},u_{\alpha'}^{\ \underline \beta})$ 
forms an element
of $Spin(1,10)$. As is clear from the notation, the indices with an
overbar take the same  range as those with  an underline. We
recall that the connection between the Lorentz and spin groups 
is given by
\bea 
u_{\overline \alpha}^{\ \underline \gamma} u_{\overline
\beta}^{\ \underline \delta}
{(\Gamma^{\underline a})}_{\underline \gamma \underline \delta }
={(\Gamma^{\overline b})}_{\overline \alpha \overline\beta }
u_{\overline b}^{\ \underline a}\ .
\label{eq:susythree}
\eea

{}For a flat target superspace the
super-reparameterisation invariance reduces to translations and rigid
supersymmetry transformations. The latter take the form
\bea
\delta x^{\underline n}= {i\over2}\Theta \Gamma^{\underline n}
\epsilon\ ,\ \
\delta \Theta ^{\underline \mu}= \epsilon^{\underline \mu}\ .
\label{eq:susyone}
\eea
Unlike other formulations, the super-embedding approach of~\cite{HSW,HS} 
is invariant under super-reparameterisations of the
worldvolume, that is, invariant under
\bea
\delta z^{ M} = -v^{ M} \ ,
\eea
where $v^{ M}$ is a supervector field on the fivebrane worldvolume.
The corresponding motion induced on the target space $\underline M$
is given by
\bea
\delta Z^{\underline B} = v^A E_A^{\ \underline B}\ ,
\eea
where $v^M= v^AE_A^{\ M}$ and rather than use the embedding
coordinates $Z^{\underline N}$ we  referred the variation to the background
tangent space, i.e.
$\delta Z^{\underline B}\equiv \delta 
Z^{\underline M} E_{\underline M}^{\ \ \underline B}$.
We are interested  in supersymmetry transformations and
so consider
$v^a=0,\ v^\alpha
\not=0$; with this choice and including the rigid supersymmetry
transformation of the target space of equation (\ref{eq:susyone}) the
transformation of $\Theta^{\underline \alpha}$ is
given by~\cite{HSW}
\bea
\delta\Theta^{\underline \alpha} = v^{\beta}
E_{\beta}^{\ \underline \alpha}
+ \epsilon^{\underline\alpha}\ .
\label{eq:vEe}
\eea
The local supersymmetry transformations $v^{\alpha}$ are used to
set $\Theta^{\alpha}=0$ which is part of the static gauge choice.
However, by combining these transformations with those of the rigid
supersymmetry  of the target space $\epsilon ^{\alpha}$ we find a
residual rigid  worldvolume supersymmetry which is determined by the
requirement  that the gauge choice $\Theta^{\alpha}=0$ is preserved.
Consequently,  we require $v^{\beta}E_{\beta}^{\ \alpha} =
-\epsilon^{\alpha}$.
{}Following the discussion in~\cite{one} 
the variation of the remaining spinor is given by
\bea
\delta\Theta^{ \alpha^\prime} = v^{\beta}
E_{\beta}^{\  \alpha^\prime}=
 v^{\beta} E_{\beta}^{\  \underline \gamma}
(E^{-1})_{\underline \gamma}^{\ \delta }E_{\delta}^{\ \alpha^\prime}\ ,
\eea
where we have set the non-linearly realized symmetry parameterized by 
$\e^{\alpha'}$ to zero. 
Introducing the projectors~\cite{HSW}
\bea
(E^{-1})_{\underline \alpha }^{\ \beta}E_\beta^{\ \underline
\gamma} ={1\over 2}(1+\Gamma)_{\underline \alpha }^{\ \underline
\gamma},\quad  
(E^{-1})_{\underline \alpha }^{\
\beta^\prime}E_{\beta^\prime}^{\ \underline \gamma} 
={1\over 2}(1-\Gamma)_{\underline \alpha }^{\ \underline \gamma}\ ,
\label{eq:pro}
\eea
we then find that the supersymmetry transformation for the
fermions is given by
\bea
\delta\Theta^{\alpha'} = -{1\over 2}\epsilon^{\gamma}
(1+\Gamma)_{\gamma }^{\ \alpha ^\prime}
+ {1\over 2}\delta\Theta^{\gamma'}
(1+\Gamma)_{\gamma^\prime }^{\ \alpha ^\prime}\ .
\eea
Hence we may write the variation of the spinor as
\bea
\delta\Theta^{\gamma'}\left({1-\Gamma\over 2}
\right)_{\gamma^\prime }^{\ \alpha ^\prime}
= -{1\over 2}\epsilon^{\gamma}(\Gamma)_{\gamma }^{\ \alpha ^\prime}\ .
\label{eq:blip}
\eea
Note that since only primed indices occur, the matrix 
${1\over 2}(1-\Gamma)_{\gamma^\prime }^{\ \alpha ^\prime}$ is
invertible. Therefore by multiplying by its inverse we find the variation
of  $\delta\Theta^{\gamma'}$. 

Bosonic configurations will preserve some supersymmetry if there exist
spinors $\epsilon$ such that $\delta\Theta^{\gamma'}$ vanishes in the limit
$\Theta^{\underline \alpha }=0$.
It will actually be more convenient to look for the conditions required
for the vanishing of the right hand side of (\ref{eq:blip}). We thus write 
(\ref{eq:blip}) as  
\bea
\hat \delta\Theta^{\alpha'}=-{1\over2}\epsilon^{\gamma}
(\Gamma)_{\gamma }^{\ \alpha ^\prime}\ ,
\eea
where we have absorbed the factor of ${1\over2}(1-\Gamma)$ into the 
definition of ${\hat \delta}\Theta^{\alpha'}$. 
To further analyse this expression we are required to find
$E_{A}^{\ \underline A}$, or equivalently the $u$'s of $SO(1,10)$ and
$Spin(1,10)$, in terms of the component fields in the limit
$\Theta^{\underline \alpha }=0$.  Using equation (\ref{eq:susytwo}), 
the Lorentz
condition $u_c^{\ \underline a} 
\eta_{\underline a \underline b} u_d^{\ \underline b}=\eta_{cd}$  and the
static gauge choice
$X^n=x^n$ we find that
\bea(u_a^{\ b},u_a^{\ b'})=(e_a^{\ n} \delta _n^b,e_a^{\ n}\partial _n
x^{b'} )\ ,
\eea
where $g_{nm}=e_n^ae_m^b\eta _{ab}
= \eta_{nm}+ \partial_nX^{a^\prime}\partial_mX^{b^\prime}\delta
_{a'b'}$. Using the remaining Lorentz conditions we find, up to a
local
$SO(5)$ rotation, that the full Lorentz matrix
$u_{\overline a}^{\ \underline a}$ is given by
\bea
u =\left(\matrix{
e^{-1}                              &e^{-1}\partial X \cr
-d^{-1}{(\partial X)}^T {(\eta_1)}^T&d^{-1} \cr }\right)\ ,
\label{eq:susysix}
\eea
where the matrix $d$ is defined by the condition
$dd^T= I+ {(\partial X)}^T \eta_1 {(\partial X)}$,
${(\partial X)}^T $ is the transpose of the matrix ${(\partial_n
X^{a'})}$  and $\eta_1$ is the Minkowski metric on the fivebrane and is
given by
$\eta_1 = diag (-1,1,1,1,1,1)$.
The $u_{\overline \alpha}^{\ \underline \beta}\in \ Spin(1,10)$ corresponding
to the above
$u_{\overline a}^{\ \underline b}\in \ SO(1,10)$ are found using
equation (\ref{eq:susythree}).

We now consider in more detail the decomposition of the spinor indices. We
recall that the bosonic  indices of the fields  on the fivebrane can
be decomposed into longitudinal and transverse indices i.e.
$\underline a= (a,a')$ according to the decomposition of the Lorentz
group
$SO(1,10)$ into
$SO(1,5)\times SO(5)$. The corresponding decomposition of the
spin group is $Spin (1,10)\to\ Spin(1,5) \times USp(4)$. The spinor
indices of the groups $Spin(1,5)$ and  $ USp(4)$ are denoted
by  $\alpha,\beta,...=1,...,4$ and $i,j,...=
1,...,4$  respectively. Six-dimensional Dirac spinor indices normally take
eight values, however the spinor indices we use for $Spin(1,5)$
correspond to  Weyl spinors. Although we began with spinor
indices $\underline \alpha $ that took  thirty-two dimensional values
and were broken into two pairs of indices each taking sixteen values
${\underline
\alpha}=(\alpha,\alpha')$, in the final six-dimensional
expressions the spinor indices  are further decomposed according
to the above decomposition of the spin groups and we take
$\alpha\rightarrow\alpha i$
and $\alpha'\rightarrow{}_{\alpha}^{i}$ when appearing as
superscripts and $\alpha\rightarrow\alpha i$
and $\alpha'\rightarrow{}^{\alpha}_{i}$ when appear as subscripts~\cite{HSW}.
It should be clear whether we mean $\alpha$ to be sixteen or
four dimensional
depending on the absence or presence of $i,j,...$ indices respectively.
{}For example, we will write  $\Theta ^{\alpha'}\to\Theta_{\alpha}^{i}$.

Using the corresponding decomposition of the spinor indices, the
eleven dimensional $\Gamma$-matrices can be written as
\bea(\Gamma^{a'})_{\underline \alpha}^{\  \underline
\beta}=(\gamma^{a'})_{i}^{\ j}
\left (\matrix{  \delta_{\alpha}^{\beta}&0\cr
0&-\delta^{\alpha}_{\beta}\cr}\right),\
(\Gamma^a)_{\underline \alpha}^{\  \underline \beta}=\delta
_{i}^{j}\left (\matrix{ 0&(\gamma^a)_{\alpha \beta}\cr
(\tilde \gamma^a)^{\alpha \beta}&0\cr}\right)\ ,
\label{eq:decomp}
\eea
where ${\gamma}^a = \gamma^0{\tilde \gamma}^a\gamma^0$.
Using this equation the eleven dimensional  $\Gamma$-matrices
with several indices can be
expressed as
\bea
(\Gamma^{a_1\ldots a_{2n} b^\prime _1\ldots
b^\prime_m})_{\underline \alpha}^{\ \  \underline \beta } &=&
(\gamma^{b^\prime _1\ldots b^\prime_m})_{i}^{\ j}
\left(\matrix{
(\gamma^{a_1\ldots a_{2n}})_\alpha ^{\ \beta} &0\cr
 0 &(-1)^m(\tilde \gamma^{a_1\ldots a_{2n}})^\alpha _{\
\beta}\cr}\right)\ ,\nonu\\
(\Gamma^{a_1\ldots a_{2n+1}b^\prime _1\ldots
b^\prime_m})_{\underline \alpha}^{\ \  \underline \beta } &=&
(\gamma^{b^\prime _1\ldots b^\prime_m})_{i}^{\ j}
\left(\matrix
{0 &(-1)^m (\gamma^{a_1\ldots a_{2n+1}})_{\alpha \beta} \cr
  (\tilde \gamma^{a_1\ldots a_{2n+1}})^{\alpha \beta} & 0\cr}\right)\ ,
\eea
where, for example, $\gamma^{a_1\ldots 
a_{2n}}\equiv\gamma^{[a_1}\tilde\gamma^{a_2}
\gamma^{a_3}\dots\tilde\gamma^{a_{2n}]}$.

We will need the relationship
\bea{(\gamma^{a_1a_2\ldots a_n})}
= -{1\over (6-n)!}(-1)^{{n(n+1)\over 2}}
\epsilon ^{a_1a_2\ldots a_na_{n+1}\ldots a_6}
\gamma_{a_{n+1}\ldots a_6}\ ,
\label{eq:susyseven}
\eea
for the chiral six dimensional $\gamma$-matrices.
The other chiral six dimensional $\tilde \gamma$-matrices
satisfy an identical condition except for an additional minus sign
on the right hand side.

Using the expressions
for the supervielbeins of equation (\ref{eq:susyfour}) 
in terms of the $SO(1,10)$
matrices, the  variation of the spinor can  be written as
\bea
\hat \delta\Theta^{\gamma'}=
 -\epsilon^{\gamma}
{(u^{-1})}_{\gamma}^{\ \beta}u_{\beta}^{\ \gamma'}
= -{1\over 2}\epsilon^{\gamma}
{(u^{-1})}_{\gamma}^{\ \underline \beta}
{(1-{1\over 6!}\epsilon^{a_1a_2a_3a_4a_5a_6}
\Gamma_{a_1a_2a_3a_4a_5a_6})}_{\underline
\beta}^{\ \underline \delta }u_{\underline \delta }^{\ \gamma'}.
\eea
The last step in the above equation used the relation
\bea
-{1\over 6!}\epsilon^{a_1a_2a_3a_4a_5a_6}
{(\Gamma_{a_1a_2a_3a_4a_5a_6})}_{\underline
\alpha}^{\ \underline \beta }= \delta_i^{\ j}
\left(\matrix{  \delta_{\alpha}^{\beta}&0\cr
0&-\delta^{\alpha}_{\beta}\cr}\right)\ .
\eea
Using equation (\ref{eq:susythree}) we find that
\bea
\hat \delta\Theta^{\gamma'} = -{1\over 2\cdot 6!}
\epsilon^{a_1a_2a_3a_4a_5a_6}
u_{a_ 1}^{\ \underline b_1} u_{a_2 }^{\ \underline b_2}
u_{a_ 3}^{\ \underline b_3} u_{a_ 4}^{\ \underline b_4}
u_{a_ 5}^{\ \underline b_5} u_{a_6 }^{\ \underline b_6}
\epsilon^{\alpha}{(\Gamma_{\ \underline b_1
\underline b_2 \underline b_3\underline
b_4 \underline b_5\underline b_6})}_{\alpha}^{\ \gamma'}\ .
\label{eq:susylast}
\eea

Equation (\ref{eq:susylast}) 
however contains an eleven dimensional $\Gamma$-matrix
that involves the upper off diagonal block and as such it vanishes
unless the $\underline b_i$ indices take values in the longitudinal
direction an odd number of times. Substituting in this matrix we find
that
\bea
\hat \delta \Theta_{\beta}^{\ j} =
&-&{1\over 2}{\rm det} (e^{-1}) \epsilon^{\alpha i}
\Big\{ \partial_a X^{c'}{(\gamma^a)}_{\alpha\beta}
{(\gamma_{c'})}_i^{\ j}\nonu\\
&&\hskip2.4cm  - {1\over 3!} \partial_{a_1} X^{c^\prime_1} \partial_{a_2}
X^{c^\prime_2}\partial_{a_3} X^{c^\prime_3}
(\gamma^{a_1a_2a_3})_{\alpha\beta}
{(\gamma_{c^\prime_1 c^\prime_2 c^\prime_3})}_i^{\ j}\nonu\\
&&\hskip2.4cm+{1\over 5!} \partial_{a_1} X^{c^\prime_1}\dots 
\partial_{a_5} X^{c^\prime_5} (\gamma^{a_1\ldots a_5})_{\alpha\beta}
{(\gamma_{c^\prime_1 \ldots c^\prime_5})}_i^{\ j}\Big\}\ .\nonu\\
\label{eq:susy}
\eea
When deriving this equation we have used equation (\ref{eq:susyseven}) and 
equation (\ref{eq:susysix}) for the $u$'s.
In the next section we will derive Bogomol'nyi equations
for bosonic configurations 
with vanishing self-dual three form
which preserve some worldvolume supersymmetry, i.e. configurations associated 
with
the vanishing of
(\ref{eq:susy}). We will do this by further manipulating (\ref{eq:susy}) by
imposing the projections on the spinor $\epsilon$ that we obtained in the
last section from considerations of orthogonally intersecting branes.  

Before proceeding to that analysis, 
it is interesting to consider the conditions for the preservation of
supersymmetry without using static gauge. Clearly
$\delta\Theta^{\underline \alpha }=0$ implies that
$v^\beta E_\beta ^{\ \underline \alpha }=-\epsilon^{\underline \alpha }$.
Multiplying by the inverse of the embedding matrix this condition
is equivalent to the two conditions
$v^\beta=-\epsilon^{\underline \alpha }
(E^{-1})_{\underline \alpha}^{\ \beta }$ and
$\epsilon^{\underline \alpha }(E^{-1})_{\underline \alpha }^{\ \beta'}
=0$.
Since $v^\beta$ is an arbitrary function the first of these equations
is automatically satisfied. The second condition is equivalent
to $\epsilon^{\underline \alpha }(E^{-1})_{\underline \alpha }^{\ \beta'}
E_{\beta'}^{\ \underline \gamma}=0$ which using the projectors
of equation  (\ref{eq:pro}) we
may rewrite as
\bea
\epsilon^{\underline \alpha }(1-\Gamma )_{\underline
\alpha}^{\ \underline \gamma}=0\ .
\label{eq:n&s}
\eea
Hence this is the  necessary and sufficient condition for the
preservation of supersymmetry.

We can now make contact with the
work of~\cite{BBS,BBMOOY}. 
{}For the static
configurations which are  studied in this paper the matrix
$\Gamma$ takes the form
\bea
\Gamma=-{1\over 5!} {\rm det} (e^{-1})
\epsilon^{m_1m_2m_3m_4m_5}
\partial_{m_1}X^{\ \underline b_1} \partial_{m_2}X^{\ \underline b_2}
\partial_{m_3}X^{\ \underline b_3} \partial_{m_4}X^{\ \underline b_4}
\partial_{m_5}X^{\ \underline b_5}
\Gamma_0\Gamma_{\ \underline b_1\underline b_2 \underline b_3\underline
b_4 \underline b_5}\ ,
\eea
where the sums exclude the value $0$. Although the matrix $\Gamma$
is in general not a hermitian matrix it is for the case of static
configurations. One can also verify that it is symmetric in
its spinor indices.

{}Following similar  arguments to those of~\cite{BBS} for the case
of the  Euclidean two brane we conclude that
\bea
\eta^\dagger(1-\Gamma)(1-\Gamma)\eta =
\eta^\dagger(1-\Gamma)\eta\ge 0\ ,
\eea
where $\eta= \epsilon^\dagger$.
The transverse coordinates will not depend on all the longitudinal
coordinates of the brane. Let us suppose that they depend on $q$
spatial coordinates leaving $p=5-q$ spatial coordinates upon which
there is no dependence. In static gauge the matrix $\Gamma$ then
further simplifies
\bea
\Gamma=-{1\over q!} {\rm det} (e^{-1})
\epsilon^{m_1\dots m_q}
\partial_{m_1}X^{\ \underline b_1} \ldots\partial_{m_q}
X^{\ \underline b_q}
\Gamma_{0\ldots p}\Gamma_{\ \underline b_1\ldots \underline b_q}
\label{eq:bound}
\eea
where ${\rm det} e$ is the determinant of the vielbein induced on
the embedded surface. Integrating equation (\ref{eq:bound}) over the $q$ 
longitudinal coordinates of the brane  we find that
\bea
\int d^{q} x ( {\rm det} e)\eta^\dagger\eta
&\ge& \int d^{q}x( {\rm det} e)\eta^\dagger \Gamma \eta\nonu\\
&=&-\int d^{q} x {1\over q!}
\epsilon^{m_1m_2m_3\dots m_q}
\partial_{m_1}X^{\ \underline b_1} \ldots\partial_{m_q}
X^{\ \underline b_q}
\eta^\dagger \Gamma_{0\ldots p}\Gamma_{\ \underline b_1\ldots
\underline b_q}\eta\ .\nonu\\
\label{eq:boundtwo}
\eea
Hence we find that the volume of the volume of the embedded surface is
greater than or equal to the integral of the  form 
$-{1\over q!}
\epsilon^{m_1m_2m_3\dots m_q}
\partial_{m_1}X^{\ \underline b_1} \ldots\partial_{m_q}
X^{\ \underline b_q}
\eta^\dagger \Gamma_{0\ldots p}\Gamma_{\ \underline b_1\ldots
\underline b_q}\eta$.
This is precisely the statement~\cite{HL} that the form calibrates
the embedded surface. {}Furthermore there is equality if and only if
supersymmetry is preserved. 

To illustrate how this works in  detail let us consider the
particular example of (\ref{eq:nine}). 
In this case four of the transverse fields of the
fivebrane are active and they depend on only four of the
longitudinal coordinates of the brane (i.e. $q=4$). Thus we have a
four dimensional space embedded in eight dimensions
which are made up of the four longitudinal coordinates of the brane
and  the four active coordinates of the fivebrane. In
this case the form of the  right hand side of ({\ref{eq:boundtwo})
has the components
\bea
-\partial_{m_1}X^{\ \underline b_1} \ldots\partial_{m_4}
X^{\ \underline b_4}
 \eta^\dagger \Gamma_{0 5}\Gamma_{\ \underline b_1\ldots
\underline b_4}\eta\ ,
\eea
where the sum over the $b_i$ excludes the values $0,5,10$. Since
$\Gamma_{0 5}\epsilon =-\epsilon$, this is just the  pull back to 
the fivebrane world surface of the four form
$\eta^\dagger \Gamma_{\ \underline b_1\ldots \underline b_4}\eta$. 
This form  lives on the eight-dimensional space and,
given the projections in (\ref{eq:nineproj}),  
is none other than the $Spin(7)$ 
invariant  self-dual four form $\Omega$ which lives on this 
eight-dimensional space (see for example \cite{jmfetal}). 
One can work out the calibrating form for all
the spaces considered in this paper in a similar manner.

{}Finally, it is interesting to compare the worldsurface supersymmetry of the 
spinor with that of $\kappa$-supersymmetry. In fact $\kappa$-supersymmetry 
is just a consequence of worldvolume supersymmetry 
which is found by taking~\cite{HSW}
$v^\beta =\kappa^{\underline \gamma}E_{\underline \gamma}^{\ \beta}$.
Making this replacement in equation (\ref{eq:vEe}) and using the projector of 
equation (\ref{eq:pro})
we find the 
standard result for the $\kappa$ transformation
\bea
\delta \Theta^{\underline \alpha } = {1\over 2}\kappa^{\underline \gamma}
(1+\Gamma)_{\underline \gamma}^{\ \underline \alpha} 
+\e^{\underline \alpha }\ .
\label{eq:kappa}
\eea
In addition setting
$\Theta ^{\alpha }=0$ in static
gauge requires ${1\over2}\kappa^{\underline \gamma}
(1+\Gamma)_{\underline\gamma}^{\ \alpha}
+\epsilon^{\alpha}=0$ and following the same argument as before
we find the variation of the remaining spinor is given by 
\bea
\delta\Theta^{\beta'}(1-\Gamma)_{\beta'}^{\ \alpha'} 
&=& {1\over 2}\kappa^{\underline \gamma}
(1+\Gamma)_{\underline\gamma}^{\ \beta}
(1+\Gamma)_{\beta}^{\ \alpha'}\nonu \\
&=&-\epsilon^{\beta}{(1+\Gamma)_\beta}^{\alpha'}\ ,
\eea 
again setting $\e^{\alpha'}=0$, which is the same as (\ref{eq:blip}).
Thus one can find
the conditions for supersymmetry preservation by studying either 
worldvolume or $\kappa$-supersymmetry.
Given that the origin of  $\kappa$-supersymmetry
is worldsurface supersymmetry this is to be expected.

\section{Geometry and Calibrations}

In section two above we wrote down static 
intersecting brane configurations which
preserve some fraction of spacetime supersymmetry. Let us now examine these
configurations from the point of view of the worldvolume of the first
fivebrane. In particular we shall further manipulate
the full non-linear supersymmetry
conditions on the worldvolume theory (\ref{eq:susy}) using the projection
operators associated with each of the configurations in section two. 
We will obtain differential equations 
for the coordinates of  
all the manifolds constructed above which correspond precisely to the
necessary and sufficient conditions of Harvey and Lawson for these to be
calibrated manifolds. 
We will see that all of these configurations correspond to the standard
K\"ahler, Special Lagrangian and exceptional calibrations of the
mathematical literature. 
As calibrated manifolds they all have minimal area in 
their homology class~\cite{HL}. 
Thus they all solve the field equations of the fivebrane
with the three form set to zero.

\subsection{K\"ahler Submanifolds}

Let us consider the case of an $n$ complex dimensional manifold
embedded in ${\bf C}^m \cong {\bf R}^{2m}$ with $m>n$. It is
helpful to introduce the complex coordinates 
\bea
z^{\mu} &=& x^{2\mu-1} + ix^{2\mu}\ , \ \ \mu = 1,2,3,...,n \nonu\\
Z^{\alpha} &=& X^{2\alpha+4} + i X^{2\alpha+5}\ , \ \ \alpha = 1,2,3,...,m-n
\eea
and their complex conjugates $z^{\bar \mu}$ and $Z^{\bar \alpha}$. Let us
denote the corresponding $\gamma$-matrices by $\gamma^{\mu}$ and 
${\gamma'}_{\alpha}={1\over2}{\gamma'}^{{\bar \alpha}}$. Here and in the 
rest of this paper we denote the transverse
$\gamma$-matrices with primes to distinguish them from the worldvolume 
$\gamma$-matrices. These furnish 
commuting representations of the $2n$-dimensional and  $2m$-dimensional 
Clifford algebras respectively;
\bea
\{\gamma^{\mu},\gamma^{\nu}\} &=& \{\gamma^{\bar \mu},\gamma^{\bar \nu}\}=0\ ,
\ \ \ \{\gamma^{\mu},\gamma^{\bar \nu}\} = 2\delta^{\mu\bar \nu}\ ,\nonu\\
\{{\gamma'_{\alpha}},{\gamma'}_{\beta}\} &=& 
\{{\gamma'}_{\bar \alpha},{\gamma'}_{\bar \beta}\}=0\ ,
\ \ \ \{{\gamma'}_{\alpha},{\gamma'}_{\bar \beta}\} = 
2\delta_{\alpha\bar \beta}\ .
\eea
We then consider the projections
\bea
\epsilon\gamma^{\mu}{\gamma'}_{\alpha} = 0 \ .
\eea
One can easily check that these form a commuting set of $n(m-n)$ projectors,
although they are not always independent. Indeed for 
$(n,m)=(1,2),(1,3),(2,3)$ one finds the configurations 
(\ref{eq:two}),(\ref{eq:three}),(\ref{eq:five})  
which preserve
$1/2,1/4,1/4$ of worldsheet supersymmetry respectively. The only other case
occurring on the fivebrane (i.e. with $n\le2$ and $m-n\le2$) is the 
configuration (\ref{eq:eight}) where $(n,m)=(2,4)$ and this
preserves $1/8$ of worldsheet supersymmetry (i.e. only three of the 
four projections are independent).

We now consider the linear term in (\ref{eq:susy})
\bea
0=\epsilon\left[\gamma^{\mu}\partial_{\mu}Z^{\alpha}{\gamma'}_{\alpha}
+ \gamma^{\bar \mu}\partial_{\bar \mu}Z^{\alpha}{\gamma'}_{\alpha}
+ c.c. \right]\ .
\eea
Clearly the first term is zero as a result of the projections and the
equation is satisfied if and only if the scalars are holomorphic functions; 
$\partial_{\bar \mu}Z^{\alpha} = 0$. {}For all the above cases with the 
exception of
$n=2,\ m=4$, the higher order terms vanish 
automatically. Thus the only
supersymmetric configurations correspond to holomorphic embeddings. 
{}For the $n=2,\ m=4$ case one finds a non-trivial third order term
coming from (\ref{eq:susy}). Vanishing of the full non-linear supersymmetry
then yields the equation
\bea
0=\e\Big\{\gamma^{\mu}\gamma'_{\bar \alpha}
\p_{\nu}Z^{\bar \beta}\Big[\delta^{\bar \alpha}_{\bar \beta}
\delta^{\nu}_{\mu}&-&{3\over2}\Big(\partial_{\mu}Z^{\gamma}
\partial^{\nu}Z_{\gamma}
\delta_{\bar \beta}^{\bar \alpha}- \partial_{\mu}Z_{\bar \beta}
\partial^{\nu}Z^{\bar \alpha}\nonu\\
&& - \delta_{\mu}^{\nu}
\partial^{\bar \rho}Z^{\gamma}\partial_{\bar \rho}Z_{\gamma}
\delta_{\bar \beta}^{\bar \alpha} + \delta_{\mu}^{\nu}
\partial^{\bar \rho} Z_{\bar \beta}\partial_{\bar \rho}Z^{\bar \alpha}
\Big)\Big] + c.c.\Big\}\ .\nonu\\
\eea
Clearly $\p_{\mu}Z^{\bar \alpha}=0$ is a solution however we
have not checked that it is the only solution. Note that  the
corresponding  complex submanifolds are 
calibrated by powers of the K\"ahler form $\omega$, 
${1\over n!}\omega^n$~\cite{HL}.

\subsection{Special Lagrangian Submanifolds}

Here we consider the case of an $n$-dimensional manifold embedded into 
${\bf R}^{2n}\equiv {\bf C}^n$.  
Let $i=1,2,3,\ldots, n$ and introduce the notation
\bea
{\gamma'}^i = {\gamma'}_{i+5} \ \ \ \ X_i = X^{i+5},
\eea
and again the two Clifford algebras $\gamma^i$ and ${\gamma'}^i$ commute.
We now consider the projections
\bea
\epsilon\gamma^{1}\gamma^{i}{\gamma'}^{1}\gamma'^{i} = \epsilon\ ,
\eea
where there is no sum over $i$. These projections in turn imply that
\bea
\e\gamma^i{\gamma'}^j =-\e \gamma^{j}{\gamma'}^i\ ,\quad i \ne j\ .
\eea
It is easy to see that these form a 
set of $n-1$ independent commuting projectors which correspond to the 
preservation 
of $2^{-(n-1)}$ of the worldvolume supersymmetry. Clearly the $n=1$ case is 
trivial and the
$n=2$ case corresponds to the $n=1,m=2$ complex case above. 

Lets us now consider the supersymmetry condition.  
{}First take $n=3$,  corresponding to the configuration (\ref{eq:four}) 
preserving $1/4$ of worldvolume supersymmetry. 
A little algebra shows that (\ref{eq:susy}) 
may be written as
\bea
0 = \sum_{i<j}\epsilon\gamma^i{\gamma'}^j
(\partial_iX_j - \partial_jX_i) + 
\epsilon\gamma^1{\gamma'}^1\left[\sum_i\partial_i X_i 
- {\rm det}(\partial X)\right]\ .
\eea
Therefore we find  from the first term that 
\bea
\partial_iX_j =\partial_jX_i\ ,
\eea
and so we take $X_j = \partial_jF$ for some $F$. The second term 
then gives
\bea
\partial^2 F={\rm det}({\rm Hess}F)\ ,
\label{eq:HLone}
\eea 
where $({\rm Hess}F)_{ij}=\partial_i\partial_j F$.

Now consider $n=4$ describing the configuration (\ref{eq:fifteen}) 
preserving $1/8$ of worldvolume supersymmetry. Here we find
\bea
0 = &&\sum_{i<j}\epsilon\gamma^i{\gamma'}^j
\left[\partial_iX_j - \partial_jX_i 
-3\partial_{[k}X^k\partial_lX^{l}\partial_{i]}X_j 
+3\partial_{[k}X^k\partial_lX^{l}\partial_{j]}X_i
\right] \nonu\\
&+&\epsilon\gamma^1{\gamma'}^1
\sum_i\left[\partial_i X_i -{\rm det}_{i|i}(\partial X)\right]\ ,\
\label{eq:fourHL}
\eea
where ${\rm det}_{i|j}(\partial X)$ is the determinant of the matrix found
by deleting the $i$th row and $j$th column of the matrix $\partial X$.
The simple condition $\p_i X_j - \p_j X_i=0$ has now become non-linear. 
Some work shows that it can be written as
\bea
0 &=&
(\p_m X _n - \p_n X_m)
\Big(\delta^m_i\delta^n_j-\delta^n_i\delta^m_j -{1\over2}
\delta^m_i \delta^n_j[(\p\cdot X)^2 -\p_l X_k \p^l X^k]
+\p^m X_j \p^n X_i\nonu\\
&-& (\p\cdot X)[\delta^m_j\p^n X_i +\delta^n_i\p^m X_j]
-\delta^m_i\p^n X_k \p^k X_j
-\delta^n_j\p^m X_k \p^k X_i
\Big)\ .
\label{eq:expand}
\eea
{}From this one readily sees that $\p_jX_i -\p_i X_j=0$ is still 
a solution (although we have not checked that it is the only solution). 
Again write $X_j = \partial_jF$ so 
that the first line in (\ref{eq:fourHL}) vanishes. The second line
then yields the equation
\bea
\partial^2 F =\sum_i {\rm det}_{i|i}({\rm Hess} F ) \ .
\label{eq:HLthree}
\eea

{}Finally we consider $n=5$ 
This describes the configuration (\ref{eq:sixteen}) preserving $1/16$
of the worldvolume supersymmetry. Here we find 
\bea
0 = &&\sum_{i<j}\epsilon\gamma^i{\gamma'}^j
\left[\partial_iX_j - \partial_jX_i 
-3\partial_{[k}X^k\partial_lX^{l}\partial_{i]}X_j  
+3\partial_{[k}X^k\partial_lX^{l}\partial_{j]}X_i
\right] \nonu\\
&+&\epsilon \gamma^1{\gamma'}^1
\left[\sum_i\partial_i X_i -\sum_{i\ne j}{\rm det}_{ij|ij}(\partial X)
+{\rm det}(\partial X)\right]\ ,\
\label{eq:fiveHL}
\eea
where ${\rm det}_{ij|kl}(\partial X)$ is the determinant of the matrix found
by deleting the $i$th and $j$th rows and $k$th and $l$th columns of the 
matrix $\partial X$.
Again equation (\ref{eq:expand}) appears and so we 
write  $X_i = \partial_iF$ and 
we arrive at the equation
\bea
\partial^2 F =  \sum_{i\ne j}{\rm det}_{ij|ij}({\rm Hess} F )  
- {\rm det}({\rm Hess}F)\ .
\label{eq:HLfive}
\eea

Equations (\ref{eq:HLone}),(\ref{eq:HLthree}) and (\ref{eq:HLfive}) above 
are precisely the necessary and sufficient conditions
derived by Harvey and Lawson~\cite{HL} for the embedded manifold in ${\bf C}^n$
to be Special Lagrangian. By definition such manifolds are 
calibrated by the form ${\rm Re}(dz^1\wedge\dots\wedge dz^n)$, where 
the $z^{\mu}$ are complex coordinates of ${\bf C}^n$.

\subsection{Exceptional Submanifolds}

We are now left with only a few of the configurations in section two 
left to analysis. As we will see, these cases correspond to
the exceptional calibrated submanifolds discussed in \cite{HL}. 
{}For these cases it will be
convenient to work with an explicit representation of
gamma matrices (\ref{eq:decomp}) using quaternions. Specifically we choose
\bea
\gamma^0 &=& \left(\matrix{1&0\cr0&1\cr}\right)\ ,\quad
\gamma^1 = \left(\matrix{0&1\cr1&0\cr}\right)\ ,\quad 
\gamma^2 = \left(\matrix{0&i\cr-i&0\cr}\right)\ ,\quad\nonu\\ 
\gamma^3 &=& \left(\matrix{0&j\cr-j&0\cr}\right)\ ,
\gamma^4 = \left(\matrix{0&k\cr-k&0\cr}\right)\ ,\quad 
\gamma^5 = \left(\matrix{-1&0\cr0&1\cr}\right)\ ,\nonu\\ 
\label{eq:basis1}
\eea
and
\bea
\gamma_6 &=& \left(\matrix{0&1\cr1&0\cr}\right)'\ ,\quad 
\gamma_7 = \left(\matrix{0&i'\cr-i'&0\cr}\right)'\ ,\quad 
\gamma_8 = \left(\matrix{0&j'\cr-j'&0\cr}\right)'\ ,\nonu\\
\gamma_9 &=& \left(\matrix{0&k'\cr-k'&0\cr}\right)'\ ,\quad 
\gamma_{10} = \left(\matrix{-1&0\cr0&1\cr}\right)'\ ,\nonu\\
\label{basis2}
\eea
where $(i,j,k)$ and $(i',j',k')$ are two commuting sets of quaternions that
can be realised as Pauli matrices.

\subsubsection{Cayley Submanifolds}

As before, the aim is now to reinterpret the spacetime configuration 
(\ref{eq:nine}) as a 
supersymmetric configuration on the first fivebrane. {}For this case four 
transverse scalars are excited and they should be functions of four 
coordinates on the fivebrane, i.e., the configurations should correspond to a 
four surface in eight dimensions. We will now show that the conditions for 
preserved supersymmetry after
imposing the projections lead to the Cayley differential equation in
\cite{HL} corresponding to Cayley submanifolds i.e. submanifolds that are
calibrated by the $Spin(7)$ invariant self-dual four-form $\Omega$. 

Before we present the derivation, 
we first note that the projections (\ref{eq:nineproj}) can be rewritten 
in the elegant
form
\beq
\epsilon {3\over 4}(\gamma_{ij} +{1\over 6} \Omega_{ijkl} \gamma^{kl}) =0\ ,
\label{twentyone}
\eeq
where we have taken the only non-zero components of $\Omega$ to be 
\bea
+1&=& \Omega_{1234}=\Omega_{6789}=\Omega_{3489}=\Omega_{2479}=
\Omega_{2378}=\Omega_{1379}=\Omega_{1267}=\Omega_{1368}=\Omega_{1469}
=\Omega_{2468}\ ,\nonu\\
-1&=& \Omega_{1289}=\Omega_{1478}=\Omega_{3467}=\Omega_{2369}\ ,\nonu\\
\label{eq:Odef}
\eea
which are the same as those in \cite{HL} after the redefinition 
${6789}\to {5678}$.
Thinking of $\epsilon$ as an $SO(8)$ spinor, (\ref{twentyone}) says that 
under the decomposition $SO(8)\to Spin(7)$, it is in fact a $Spin(7)$ singlet.
To see this note that the adjoint of $SO(8)$ decomposes as 
${\bf 28}\to {\bf 21} + {\bf 7}$,
where ${\bf 21}$ is the adjoint of $Spin(7)$, and that the matrix that appears
in  (\ref{twentyone}) is precisely the operator that projects onto
the ${\bf 21}$ (see, for example \cite{bobbyetal,jmfetal}). 
We thus conclude that the projection operators (\ref{eq:nineproj}) that we
obtained from considerations of orthogonally intersecting branes are 
equivalent to the more abstract 
statement that we are working with a spinor that is a $Spin(7)$ singlet.
One implication of this observation is that we expect the
same projections to appear for just two fivebranes rotated by a $Spin(7)$
rotation \cite{OT}.

Let us now begin the derivation of the Cayley equation.
We first rewrite the 
projections
using the explicit basis (\ref{eq:basis1}),(\ref{basis2}). We 
conclude the following:
\bea
\e\left(\matrix{1&0\cr0&0\cr}\right)&=&0\ , \quad
\e\left(\matrix{1&0\cr0&0\cr}\right)'=0\ ,\nonu\\
\e\left(\matrix{0&0\cr0&i\cr}\right)&=&-\e\left(\matrix{0&0\cr0&i'\cr}\right)'
\ ,\nonu\\
\e\left(\matrix{0&0\cr0&j\cr}\right)&=&-\e\left(\matrix{0&0\cr0&j'\cr}\right)'
\ ,\nonu\\
\e\left(\matrix{0&0\cr0&k\cr}\right)&=&-\e\left(\matrix{0&0\cr0&k'\cr}\right)'
\ ,\nonu\\
\label{conds}
\eea
which allows one to trade $Spin(5,1)$ matrices
for $Spin(5)$ matrices when acting on the spinor $\e$.
The signs are necessary and essentially arise from the fact that
the $Spin(5,1)$ $\gamma$-matrices commute with the $Spin(5)$ matrices.

We look for configurations with $\p_0=\p_5=0$ and all transverse scalars 
 excited except $X^{10}$. It is convenient to introduce the quaternion 
valued fields and derivatives
\bea
X'&=&X^6+i'X^7+j'X^8+k'X^9\ ,\nonu\\
\p&=&\p_1+i\p_2+j\p_3+k\p_4\ ,\nonu\\
\overline{\p}&=&\p_1-i\p_2-j\p_3-k\p_4\ .\nonu\\
\eea
We first consider the terms in the supersymmetry variation that are
linear in $X$:
\bea
\e X^{b'}\gamma_{b'}{\overleftarrow \p}_a\gamma^a&=&
\e\left(\matrix{0&X'\cr\overline{X}'&0\cr}\right)'
\left(\matrix{0&\p\cr\overline{\p}&0\cr}\right)\ ,\nonu\\
&=&\e\left(\matrix{0&0\cr0&\overline{X}'\cr}\right)'
\left(\matrix{0&0\cr0&\overline{\p}\cr}\right)
\left(\matrix{0&1\cr1&0\cr}\right)'
\left(\matrix{0&1\cr1&0\cr}\right)\ ,\nonu\\
&=&\e\left(\matrix{0&0\cr0&X\overline{\p}\cr}\right)
\left(\matrix{0&1\cr1&0\cr}\right)'\left(\matrix{0&1\cr1&0\cr}\right)\ ,\nonu\\
\eea
where $X=X^6+iX^7+jX^8+kX^9$, 
$X \overline\p\equiv \p_1 X-\p_2 Xi-\p_3 Xj-\p_4 Xk$ and we have used 
(\ref{conds}).

Next we turn to the terms in the supersymmetry variation that are cubic in
$X$. By performing similar steps we obtain
\bea
&&-{1\over 3!}\partial_{a_1} X^{b^\prime_1} \partial_{a_2}
X^{b^\prime_2}\partial_{a_3} X^{b^\prime_3}
(\gamma^{a_1a_2a_3})
{(\gamma_{b^\prime_1 b^\prime_2 b^\prime_3})}\nonu\\
&=&-{1\over 3!}\e\left(\matrix{
0&0\cr 0&\p_{a_1} \overline{X}'\p_{a_2}
X'\p_{a_3}\overline{X}'\cr}\right)'
\left(\matrix{0&1\cr1&0\cr}\right)' \gamma^{a_1 a_2 a_3}\ ,\nonu\\
&=&+{1\over 3!}\e\left(\matrix{0&0\cr 0&
\p_{[a_1}X\p_{a_2}\overline{X}
\p_{a_3]}X\cr}\right)
\left(\matrix{0&1\cr1&0\cr}\right)' \gamma^{a_1 a_2 a_3}\ ,\nonu\\
&=&+{1\over 3!}\e\left(\matrix{
&0\cr 0&\p_{a_1} X\times 
\p_{a_2}{ X}
\times \p_{a_3}X\cr}\right)
\left(\matrix{0&1\cr 1&0\cr}\right)'\gamma^{a_1 a_2 a_3}\ ,\nonu\\
\eea
where we have introduced the triple $\times $ product of quaternions
defined by
\beq
x\times y\times z ={1\over 2}(x\bar y z-z\bar y x)\ ,
\eeq
and we have used the fact that it is alternating. 
Next we let the indices $a_1 a_2 a_3$ run over the values
$1,2,3,4$ and substitute their explicit form using 
(\ref{eq:basis1}),(\ref{basis2}).
Combining with the terms linear in $X$ we conclude that the condition for
preserved supersymmetry is encapsulated by the differential
equation
\bea
\p_1  X-\p_2 Xi-\p_3  Xj-\p_4 Xk
&=&
\p_{2} X\times \p_{3}{X}
\times \p_{4}X
+\p_{1} X\times \p_{3}{X}
\times \p_{4}X i\nonu\\
&-&\p_{1} X\times \p_{2}{X}
\times \p_{4}X j
+\p_{1} X\times \p_{2}{X}
\times \p_{3}X k
\ . \nonu\\
\eea
This is the Cayley equation derived in \cite{HL} for submanifolds that are
calibrated by the Cayley calibration.

\subsubsection{Associative Submanifolds}
\def\vphi{\varphi}

Next consider the configuration (\ref{eq:six}) preserving $1/16$ of the 
spacetime supersymmetry.  
{}For this case four transverse
scalars are excited and they should be functions of three coordinates on 
the fivebrane.
i.e. the configurations should correspond to a three surface in seven 
dimensions.
We will now show that the conditions for preserved supersymmetry after
imposing the projections (\ref{eq:sixsusy}) lead 
to the associator equation in~\cite{HL} for associative 
submanifolds i.e. submanifolds calibrated by
the $G_2$ invariant three form $\vphi$. 
The non-zero components of $\vphi$
can taken to be
\bea 
+1&=& \vphi_{234}=\vphi_{267}=\vphi_{469}=\vphi_{379}=
\vphi_{368}\ ,\nonu\\
-1&=& \vphi_{289}=\vphi_{478}\ . \nonu\\
\eea

As in the Cayley case, we first note that our projections (\ref{eq:sixsusy}) 
can be recast in 
the 
form
\beq
\epsilon {2\over 3}(\Gamma_{ij} +{1\over 4} \psi_{ijkl} \Gamma^{kl}) =0\ ,
\label{fourteen}
\eeq
where the four-form $\psi$ is the Hodge-dual of $\vphi$ 
in the directions $\{2,3,4,6,7,8,9\}$. Specifically the non-zero components 
of $\psi$ are
\bea
+1&=& \psi_{6789}=\psi_{3489}=\psi_{2479}=
\psi_{2378}=\psi_{2468}\ ,\nonu\\
-1&=& \psi_{3467}=\psi_{2369}\ ,\nonu\\
\label{psicomp}
\eea
which are simply the components of $\Omega$ in (\ref{eq:Odef}) 
without a 1 component.
Thinking of $\epsilon$ as a $Spin(7)$ spinor, (\ref{fourteen}) says that
it is actually a $G_2$ singlet under the decomposition $Spin(7)\to G_2$.
This is because the adjoint of $Spin(7)$ decomposes as ${\bf 21} \to {\bf 
14}+{\bf 7}$
and the matrix appearing  in (\ref{fourteen}) projects onto the ${\bf 14}$, 
the  adjoint of $G_2$.

{}For this case, after we 
rewrite the projections (\ref{eq:sixsusy})
using our explicit basis (\ref{eq:basis1}),(\ref{basis2}), we 
conclude the 
following:
\bea
\e\left(\matrix{1&0\cr0&0\cr}\right)'&=&0\ , \nonu\\
\e\left(\matrix{i&0\cr 0&i\cr}\right)
&=&-\e\left(\matrix{0&0\cr0&i'\cr}\right)'\ ,\nonu\\
\e\left(\matrix{j&0\cr0&j\cr}\right)
&=&-\e\left(\matrix{0&0\cr0&j'\cr}\right)'\ ,\nonu\\
\e\left(\matrix{k&0\cr0&k\cr}\right)
&=&-\e\left(\matrix{0&0\cr0&k'\cr}\right)'\ .\nonu\\
\eea

Now we turn to the supersymmetry variation. We now define 
$X'=X^6+i'X^7+j'X^8+k'X^9$
and $\p=+i\p_2+j\p_3+k\p_4=-\overline \p$.
The terms linear in
$X$ can now be processed as follows:
\bea
\e X^{b'}\gamma_{b'} \p_a\gamma^a&=&
\e\left(\matrix{\overline{\p}&0\cr 0&\overline{\p}\cr}\right)
\left(\matrix{0&0\cr0&\overline{X}'\cr}\right)'
\left(\matrix{0&-1\cr1&0\cr}\right)\left(\matrix{0&1\cr1&0\cr}\right)'\ ,\nonu\\
&=&\e\left(\matrix{0&0\cr0&\p'\overline{X}'\cr}\right)
\left(\matrix{0&-1\cr1&0\cr}\right)\left(\matrix{0&1\cr1&0\cr}\right)'\ .\nonu\\
\eea
Similarly, the cubic terms can be rewritten
\bea
&&-\e\left(\matrix{
0&0\cr 0&\p_{[2} \overline{X}'\p_{3}
X'\p_{4]}\overline{X}'\cr}\right)'
\left(\matrix{0&1\cr1&0\cr}\right)' \gamma^{234}\nonu\\
&&=\e\left(\matrix{0&0\cr 0&\p_{2} \overline{X}'\times \p_{3}
\overline{X}'\times \p_{4}\overline{X}'\cr}\right)'
\left(\matrix{0&1\cr1&0\cr}\right)' \left(\matrix{0&-1\cr1&0}\right) .\nonu\\
\eea
After taking the hermitian conjugate 
the condition for unbroken supersymmetry is given by the differential
equation (after dropping the primes)
\beq
-\p_2 X i - \p_3 X j -\p_4 X k =  \p_2 X\times \p_3 X \times \p_4 X\ ,
\eeq
which is the associator equation that appears in \cite{HL}. Recently 
solutions to this equation have been studied  in relation to domain 
walls in MQCD~\cite{Volovich}.

\subsubsection{Coassociative Submanifolds}

Next consider the configuration (\ref{eq:seven}) preserving $1/16$ of the 
spacetime supersymmetry.  
{}For this case three transverse
scalars are excited and they should be functions of four coordinates on 
the fivebrane.
i.e. the configurations should correspond to a four surface in seven 
dimensions.
We will now show that the conditions for preserved supersymmetry after
imposing the projections (\ref{eq:sevensusy}) lead to the 
coassociator differential equation in
\cite{HL} for coassociative submanifolds i.e. manifolds calibrated by the $G_2$
invariant four-form $\psi$ that is Hodge dual to the three-form $\vphi$ in the 
associative 
case. 
{}For this case the projectors can be recast in the form 
(\ref{fourteen}) where the components of $\psi$ are now
given by (\ref{psicomp}) after the relabelling $\{2346789 \}\to \{7891234\}$.
Thus, the projections for this case also imply that the spinor is 
a $G_2$ singlet. 

To obtain the corresponding differential equation we begin by 
rewriting the projections using the explicit basis 
(\ref{eq:basis1}),(\ref{basis2}), to conclude: 
\bea
\e\left(\matrix{1&0\cr0&0\cr}\right)&=&0\ , \nonu\\
\e\left(\matrix{i'&0\cr 0&i'\cr}\right)'
&=&-\e\left(\matrix{0&0\cr0&i\cr}\right)\ ,\nonu\\
\e\left(\matrix{j'&0\cr0&j'\cr}\right)'
&=&-\e\left(\matrix{0&0\cr0&j\cr}\right)\ ,\nonu\\
\e\left(\matrix{k'&0\cr0&k'\cr}\right)'
&=&-\e\left(\matrix{0&0\cr0&k\cr}\right)\ .\\
\eea

We now have $X=i'X^7+j'X^8+k'X^9=-\overline X$
and $\p=\p_1+\p_2i+\p_3j+\p_4k$.
The  terms in the supersymmetry variation that are linear in $X$ can be 
reexpressed
\bea
\e X^{b'}\gamma_{b'} \p_a\gamma^a&=&
\e\left(\matrix{{X}'&0\cr0&{X}'\cr}\right)'
\left(\matrix{0&0\cr 0&\overline{\p}\cr}\right)
\left(\matrix{0&1\cr-1&0\cr}\right)'\left(\matrix{0&1\cr1&0\cr}\right)\ ,\nonu\\
&=&\e\left(\matrix{0&0\cr0&-X\overline{\p}\cr}\right)
\left(\matrix{0&1\cr-1&0\cr}\right)'\left(\matrix{0&1\cr1&0\cr}\right)\ .
\eea
Similarly, the cubic terms can be rewritten
\bea
&&-\e\left(
\matrix{0&0\cr 0&\overline{\p} X^{[7}{\p} X^8 \overline{\p} X^{9]}\cr}\right)
\left(\matrix{0&1\cr1&0\cr}\right)\gamma_{789}\nonu\\
&=&-\e\left(\matrix{0&0\cr 0&\overline{\p} X^{7}\times \overline{\p} X^8 
\times\overline{\p} X^{9}\cr}\right)
\left(\matrix{0&1\cr1&0\cr}\right)
\left(\matrix{0&1\cr-1&0\cr}\right)'\ .
\eea
After taking the hermitian conjugate 
the condition for unbroken supersymmetry is then given by the differential
equation
\beq
-\p X^7 i - \p X^8 j -\p X^9 k =  \p X^7\times \p X^8 \times \p X^9\ ,
\eeq
which is the coassociator equation that appears in \cite{HL}.

\section{Conclusion}

In this paper we have analysed the conditions necessary for
the fivebrane worldvolume theory to preserve some supersymmetry,
when the self-dual three form is set to zero.
Our approach was to first consider spacetime configurations
of orthogonally intersecting fivebranes in order to derive
a set of projection operators acting on the worldvolume supersymmetry
parameters. By manipulating the supersymmetry variation using
these projections we derived a set of differential equations  
for the transverse scalar fields of the worldvolume. 
These Bogomol'nyi equations are none other than the equations resulting from 
calibrated
geometries~\cite{HL}. It would be interesting to find explicit solutions
to these equations, i.e. calibrated geometries, for simple cases such
as the orthogonal brane configurations of section two, or branes in flat space 
rotated by elements of groups associated with special holonomy 
\cite{BDL,GGPT,OT} (since we might expect that these
configurations are associated with the same projection 
operators as for the orthogonal
configurations that we explicitly considered in section two).

It will be very interesting to extend the analysis of this paper to include
membranes by allowing for a non-zero self-dual three form. 
We expect that the resulting differential equations will be 
associated with a generalised notion of calibrated geometries~\cite{US}.
We pointed out in section two some
configurations  allow for pp-waves and membranes
to be introduced without breaking any addition supersymmetry. One may
expect that this has a simple interpretation in the resulting 
generalised calibrations.

In this paper we have studied fivebranes in a flat target space.
It is well known that calibrated geometries can be defined in
manifolds with special holonomy. {}For example, in eight dimensions
it is well known that in a curved manifold with reduced $Spin(7)$ holonomy 
the self-dual four form $\Omega$ is globally defined and this allows one to 
have submanifolds calibrated by $\Omega$. With a flat target space we saw that
the preserved supersymmetries for the calibrated geometries are 
$Spin(7)$ invariant spinors. This also has a natural generalisation
since manifolds with $Spin(7)$ holonomy contain parallel spinors. In a similar
manner the associative and coassociative cases will generalise to seven 
dimensional
manifolds with $G_2$ holonomy while the K\"ahler  and special Lagrangian cases 
can be generalised
to manifolds with $SU(n)$ holonomy.

It would interesting to generalise our analysis to find the
analogues of the differential equations of~\cite{HL} in a curved
manifold of special holonomy. We leave this to future work, but
we would like to mention how some of the analysis of section three
could be generalised to a curved target space.  
Examining equation (\ref{eq:susytwo}) we find that now
\bea
u_a^{\underline b}= e_a^m\partial_m X^{\underline N}
E_{\underline N}^{\underline b}\ ,
\eea
where $e_a^m g_{nm} e_d^m= \eta_{ad}$ and
$g_{nm}= \partial _n X^{\underline N}
E_{\underline N}^{\underline b}\eta _{\underline b\underline d}
\partial _m X^{\underline R}E_{\underline R}^{\underline d}$.
The equation for the matrix $\Gamma$ is still given by
\bea
-{1\over 6!}\epsilon^{a_1a_2a_3a_4a_5a_6}
u_{a_ 1}^{\ \underline b_1} u_{a_2 }^{\ \underline b_2}
u_{a_ 3}^{\ \underline b_3} u_{a_ 4}^{\ \underline b_4}
u_{a_ 5}^{\ \underline b_5} u_{a_6 }^{\ \underline b_6}
\epsilon^{\alpha}{(\Gamma_{\ \underline b_1
\underline b_2 \underline b_3\underline
b_4 \underline b_5\underline b_6})}_{\alpha}^{\ \gamma'}\ .
\eea
But if we define
$v_c^{\underline b}$ by $u_a^{\underline b}= e_a^{m}(f^{-1})_m^c
v_c^{\underline b}$ where $(f^{-1})_m^c$ is the matrix
$(f^{-1})_m^c= e_m^b u_b^c$ then we may write $\Gamma$ as
\bea
-{1\over 6!} ({\rm det}(ef)^{-1})
\epsilon^{a_1a_2a_3a_4a_5a_6}
v_{a_ 1}^{\ \underline b_1} v_{a_2 }^{\ \underline b_2}
v_{a_ 3}^{\ \underline b_3} v_{a_ 4}^{\ \underline b_4}
v_{a_ 5}^{\ \underline b_5} v_{a_6 }^{\ \underline b_6}
\epsilon^{\alpha}{(\Gamma_{\ \underline b_1
\underline b_2 \underline b_3\underline
b_4 \underline b_5\underline b_6})}_{\alpha}^{\ \gamma'}\ .
\eea
The net effect of these changes is just to make the replacement
\bea
\delta^n_c\partial_n X^{b'}\to  (f^{-1})_m^c (E_m^{b'}
+\partial_MX^{n'}E_{n'}^{b'})\ ,
\eea
in all formulae for the supervariation of the spinor.


\section*{Acknowledgments}

We would like to thank B. Acharya, F. Dowker, J. Figueroa O'Farrill, 
G. Gibbons and G. 
Papadopoulos for useful discussions.  JPG is supported in part by EPSRC.

While this paper was being prepared we learnt of the work~\cite{GP} which has
some overlap with this work.

\end{document}